\documentclass[12pt,preprint]{aastex}
\usepackage{epstopdf}

\usepackage{mathtools}

\begin{document}

\title{The Hetu'u Global Network: Measuring the Distance to the Sun Using the June 5th/6th Transit of Venus } 

\author{Jacqueline K. Faherty\altaffilmark{1}, David R. Rodriguez\altaffilmark{1}, Scott T. Miller\altaffilmark{2}}

\altaffiltext{1}{Departamento de Astronom{\'i}a, Universidad de Chile, Casilla 36-D, Santiago, Chile}
\altaffiltext{2}{Sam Houston State University}

\begin{abstract}
In the spirit of historic astronomical endeavors, we invited school groups across the globe to collaborate in a solar distance measurement using the rare June 5/6th transit of Venus.  In total, we recruited 19 school groups spread over 6 continents and 10 countries to participate in our Hetu'u Global Network.  Applying the methods of French astronomer Joseph-Nicolas Delisle, we used individual second and third Venus-Sun contact times to calculate the distance to the Sun.  Ten of the sites in our network had amiable weather; 8 of which measured second contact and 5 of which measured third contact leading to consistent solar distance measurements of $152\pm30$ million km and $163\pm30$ million km respectively. The distance to the Sun at the time of the transit was 152.25 million km; therefore, our measurements are also consistent within 1$\sigma$ of the known value.  The goal of our international school group network was to inspire the next generation of scientists using the excitement and accessibility of a rare astronomical event.  In the process, we connected hundreds of participating students representing a diverse, multi-cultural group with differing political, economic, and racial backgrounds.
\end{abstract} 


\section{Introduction}

The average distance between the Earth and the Sun, also called the astronomical unit (AU), is the fundamental constant in our solar system.  A precise measurement of the AU is required to fully utilize the three laws of planetary motion established by Johannes Kepler in the early 17th century.  KeplerÕs laws relate the spacing of the planets to one another allowing measurements to and between each planet. However, in order to establish the absolute scale of the solar system, the AU is required as a yardstick.  Consequently, for much of the 18th and 19th centuries, a primary goal in astronomy was to make a precise measurement of the AU. One of the most feasible methods proposed was to use the transit of Venus.  

The fact that planets transit, or pass directly in between Earth and the Sun, was known since Kepler determined the ordering of the solar system.  Kepler himself predicted the 1631 transit of Venus in 1626 in a paper called ``Admonitiuncula ad Curiosos Rerum Coelestium,'' but was unable to view it because the Sun had set in Europe by the time of first contact. The first known observations of a Venus transit were done by Jeremiah Horrocks and William Crabtree in 1639 (Forbes 1874; Teets 2003).

In 1716, in a paper titled ``A New Method of Determining the Parallax of the Sun,'' Edmond Halley (of comet fame) proposed a method for using the transit of Venus to determine the distance to the Sun (Forbes 1874).   As a result, by the time the subsequent transit of Venus occurred in 1761, scientists were prepared to use the event to make the measurement. 

Halley proposed to use the standard astronomical method of parallax to calculate the AU. Just as nearby objects appear to shift location relative to background objects when viewed from different perspectives, Halley surmised that Venus, being closer to Earth than the Sun, would appear to transit across a different segment of the Sun when viewed from one location on Earth than another.  Unfortunately Halley died before the 1761 transit, but based on his work, the transit of 1761 proved to be a collaborative milestone, as over 120 observers at more than 62 locations worked together to time the transit event in order to calculate the astronomical unit.

Astronomer James Short, best known as one of the most skilled telescope makers of his time, not only observed the 1761 transit, but also derived the calculations used to compute the Earth-Sun distance (Short 1761; Teets 2003).  ShortÕs calculations were based on the idea that observers at different locations would observe the Sun crossing along different chords across the surface of the Sun.  By measuring the angular difference between the chords along with the linear separation between observers that were strategically placed across the globe for the 1761 transit, he was able to calculate the AU.  Unfortunately, due to imprecise measurements of the longitudes of the observers, ShortÕs calculations provided a wide range in distances, depending on which data he used.  Fortunately, another Venus transit occurred only 8 years later in 1769, and based on previous experience from the 1761 transit, astronomers were able to calculate a more precise value for the astronomical unit of 150,839,000 km (only a 0.8\% error compared to the accepted value of 149,598,000 km; Teets 2003)

\begin{figure}[htb]
\begin{center}
\includegraphics[width=10cm,angle=0]{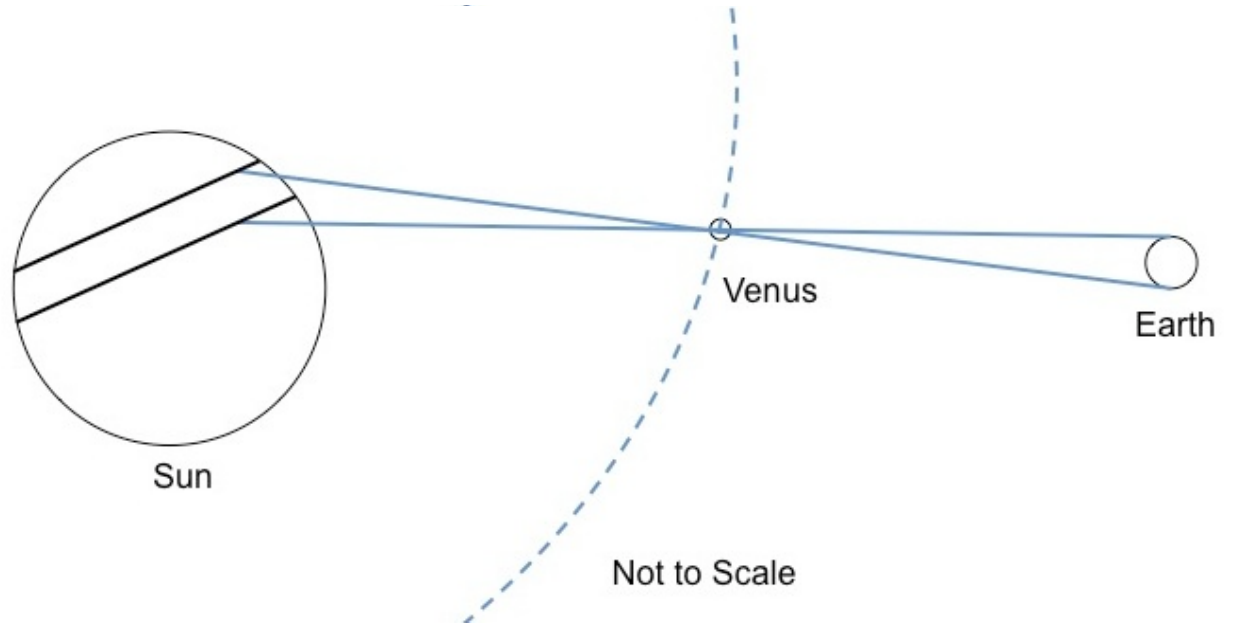}
\end{center}
\caption{As Venus transits across the surface of the Sun, observers in the northern hemisphere would observe Venus transit along a lower projection than observers in the southern hemisphere.
}
\end{figure}

While the transit of Venus provides an excellent opportunity to measure the distance from Earth to the Sun, the disadvantage is that it occurs infrequently.  In fact, the transit of Venus occurs only four times every 243 years.  The reason why it does not occur more frequently is because VenusÕs orbit is inclined relative to EarthÕs orbit. 

Venus lies closer to the Sun than Earth, and therefore completes one orbit in less time (once every 224.70 days, compared to EarthÕs period of 365.25 days).  As a result Venus and Earth line up (ie, Venus is at inferior conjunction) once every 583.92 days, which is roughly equal to 1.6 Earth years (one synodic period).  If VenusÕs orbit were not inclined to EarthÕs, then transits would happen almost every 1.6 years.  Due to VenusÕs 3.4-degree inclination to ecliptic, though, most of the time it does not pass directly between Earth and the Sun, but a little above or below.  Only when Venus is at inferior conjunction and crosses the ecliptic does a transit occur. 

\begin{figure}[htb]
\begin{center}
\includegraphics[width=10cm,angle=0]{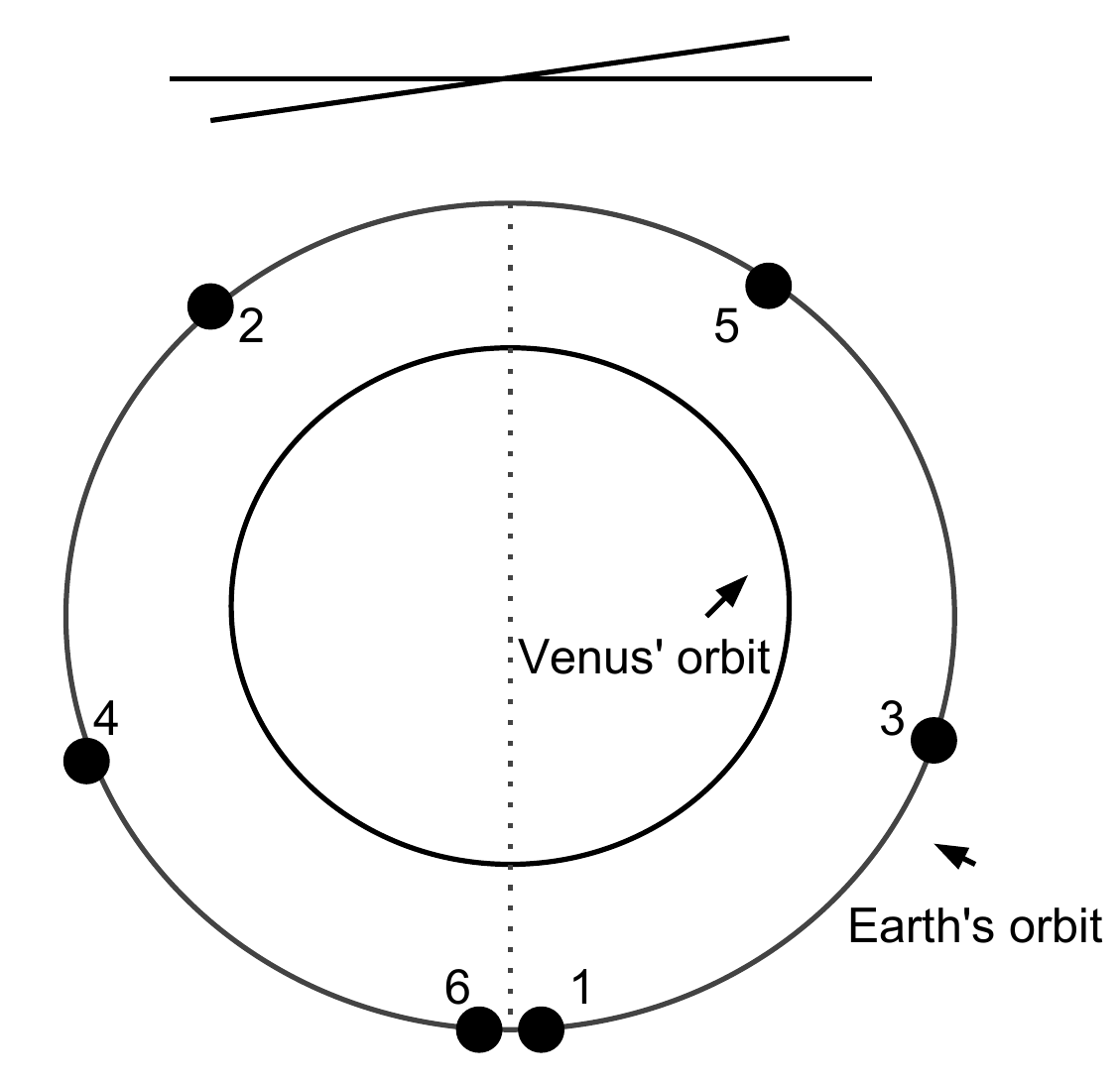}
\end{center}
\caption{A face-on view (bottom) of EarthÕs and VenusÕs orbits and points of inferior conjunction, along with a side view (top) of their orbits which demonstrates the 3.4-degree inclination of VenusÕs orbit.  The dotted line denotes the line of nodes, where VenusÕs orbit crosses the ecliptic plane.  If inferior conjunction occurs along the line of nodes, a transit will occur.
}
\end{figure}

Figure~2 demonstrates the points along EarthÕs orbit where, once every 1.6 years, Venus is at inferior conjunction.  For most of the conjunctions, this occurs when Venus is either above or below the ecliptic plane (points 2 -- 5).  It is only when Venus is at point 1 when a transit occurs.  If VenusÕs synodic period were exactly 1.6 years, every fifth inferior conjunction would occur exactly 8 years apart.  If this were the case, when one transit occurs, the next transit would occur exactly 8 years later.

However, VenusÕs synodic period is not exactly 1.6 years, but rather slightly less.  As a result, the locations of inferior conjunction shown in figure~2 precess over time.  While the rate of precession is small enough that two consecutive transits can occur eight years apart, by the time another eight years has passed, the point of inferior conjunction has moved far enough that Venus is no longer on the ecliptic plane.  It takes over 100 years before another inferior conjunction will occur at the same time that Venus passes through the ecliptic.

\section{Objectives}

Using the transit of Venus to measure the distance to the Sun is a global effort.  One must make a measurement of the transit contact points from various locations on the globe, ideally with a baseline as large as the radius of the Earth. Historically, the transit of Venus is one of the first instances of global scientific collaboration since participation by multiple countries spread across the globe is required to make the measurement feasible. For the 18th and 19th century transits, efforts brought together scientists, engineers, and laymen from multiple continents all belonging to countries with differing political, economic, and cultural backgrounds. At the time, the efforts of this global network were at the cutting edge of science. However, in modern-day astronomy, the distance to the Sun is known to a fraction of a km and has been measured numerous ways with advanced modern technology. Therefore, the appeal of the 2012 transit of Venus was not to rival the most advanced and precise techniques of our time.  Rather, in the spirit of the scientific collaborations of the 18th and 19th centuries, we recognized this transit as a unique opportunity for global participation and one that could be used to join school-aged students together to inspire our next generation of scientists with a truly remarkable and rare astronomical event.  Our goal was to link together numerous school groups from different continents and countries strategically placed across the globe so timing measurements would allow us to collaborate on a solar distance measurement.  

\section{Calculating the Distance to the Sun}

While the method originally presented by Halley requires a full observation of the transit from more than one location, the method we used is more simple in that it only requires the observers to note the time of ingress interior (second contact, or the moment when Venus first completely blocks a portion of the Sun) or egress interior (third contact, or the moment when Venus last completely blocks a portion of the Sun) along with the observerÕs latitude and longitude. This alternate method was initially developed by the French astronomer Joseph-Nicolas Delisle for the 1753 transit of Mercury. While perhaps not as well known as HalleyÕs method, it is applicable for a greater part of the Earth given that it does not require an observer to witness the full transit.

\begin{figure}[htb]
\begin{center}
\includegraphics[width=10cm,angle=0]{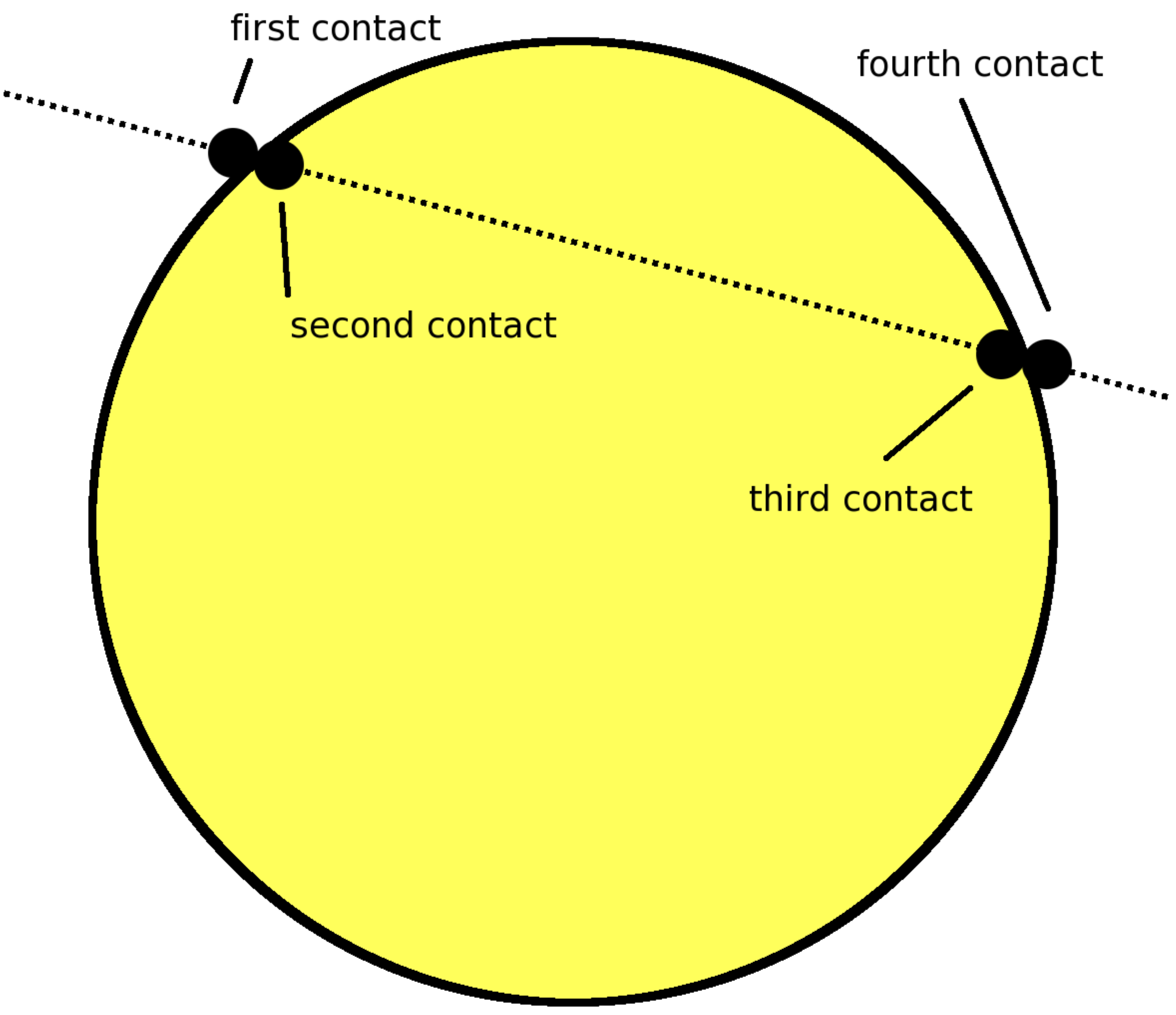}
\end{center}
\caption{Schematic diagram of the four contact points of Venus, as it transits in front of the Sun.
}
\end{figure}

Figure~3 illustrates the path of Venus in the equatorial view as it transits across the Sun.  The four contact points of note are:  ingress exterior (first contact), when Venus just touches the disk of the Sun; ingress interior (second contact), when Venus is fully inside the solar disk; egress interior (third contact), when Venus is just about to leave the disk; and egress exterior (fourth contact), when Venus has just completely left the disk of the Sun.

Consider the motion of the Earth as it moves into the shadow of Venus as seen from the Sun. At ingress exterior, the Earth is just touching the shadow and we see Venus just hitting the Sun's disk. At ingress interior, the Earth has fully moved into the shadow and the transit is well underway.  In that time the Earth has moved one full Earth diameter (2 Earth radii: $2R_E$).  As Earth moves one Earth diameter around the Sun in its orbit, it subtends an angle $\theta$, such that:
\begin{eqnarray}
2 R_E = D \theta
\end{eqnarray}
where D is the distance between Earth and the Sun.  Because the angle $\theta$ can be troublesome to measure, we can rewrite it as the amount of time it take the Earth to travel at a given angular speed:
\begin{eqnarray}
2 R_E = D \omega \Delta t 
\end{eqnarray}
\begin{eqnarray*}
\text{or:}\, D = \frac{2R_E}{\omega \Delta t}
\end{eqnarray*}

Therefore, if we can measure the time $\Delta t$ between ingress exterior and ingress interior, we can estimate the distance to the Sun since we already know the radius of the Earth and EarthÕs angular speed.
 
This assumes, though, that Venus is motionless while the Earth revolves, which is not the case.  Therefore, instead of EarthÕs angular speed, we need to use the relative speed between Venus and Earth.  In order to determine this, we can assume circular orbits for both planets (which, while not true, is a close enough approximation for the purpose of this calculation).  Given the period of revolution, P, of both planets, the angular speed for a planet is equal to $\omega = 2\pi/P$.  For Venus, this is equal to $3.2364 \times 10^{-7}$ /second, and for Earth, $1.992 \times 10^{-7}$ /second.  Thus, the relative angular motion is equal to $1.244 \times 10^{-7}$ /second.
 
This calculation would be complete if Earth enters the shadow of Venus directly through the center, but it does not and therefore we need to account for this difference.  To do so we must measure the `impact parameter' ($p$) of the path of the Earth through Venus's shadow:

\begin{figure}[htb]
\begin{center}
\includegraphics[width=10cm,angle=0]{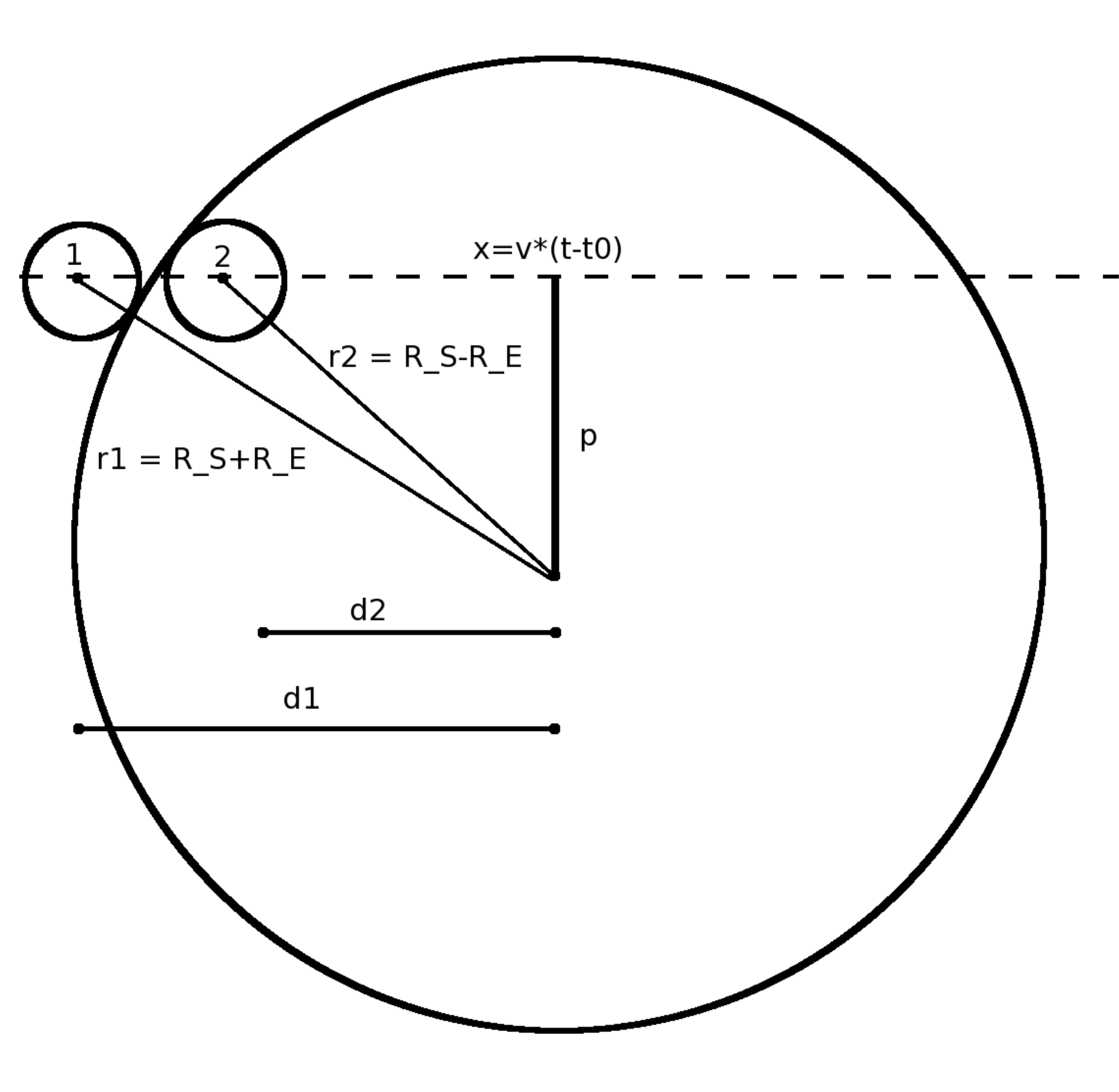}
\end{center}
\caption{A measurement of the impact parameter ($p$) as Earth (the small circle) enters into the shadow of Venus (the large circle).  The points of first and second contact are shown.
}
\end{figure}
 
The variable $r$ corresponds to the true separation between the center of the shadow and the center of the Earth while $d$ corresponds to the horizontal projection of this distance. The variable $p$ is the vertical separation between the path of the Earth and the center of Venus's shadow. Since these are right triangles, the following holds:
\begin{eqnarray}
r^2 = p^2 + d^2
\end{eqnarray}

The distance $r_1$ is simply the sum of the radius of the shadow ($R_S$) and the radius of the Earth ($R_E$), whereas $r_2$ is the difference between these two quantities.
Also, recall from above that when the Earth fully enters the shadow it has moved a horizontal distance $2R_E$.  This can be expressed more generally as:
\begin{eqnarray}
d_1 - d_2 = v \Delta t'
\end{eqnarray}
where $v$ is the linear velocity ($v = \omega D$) of Earth and $\Delta t'$ is the time between event 1 and 2.  We can combine the relationships as follows:
\begin{eqnarray*}
d_1 - d_2 = v \Delta t' \\
d_1 + d_2 = 2 \sqrt{R_S^2-p^2} \\
r_1-r_2 = 2 R_S \\
r_1+r_2 = 2 R_E
\end{eqnarray*}

Using these equations, we can recombine them to derive the following:
\begin{eqnarray*}
d_1^2 = r_1^2 - p^2 \\
d_2^2 = r_2^2 - p^2 \\
d_1^2 - d_2^2 = r_1^2 - r_2^2 \\
(d_1-d_2) (d_1+d_2) = (r_1-r_2) (r_1+r_2) \\
(v \Delta t') 2 \sqrt{R_S^2-p^2} = (2 R_S) (2 R_E) \\
\omega D \Delta t' \sqrt{R_S^2-p^2} = 2 R_S R_E
\end{eqnarray*}
\begin{eqnarray}
D = \frac{2 R_S R_E}{\omega \Delta t' \sqrt{R_S^2-p^2}} = \frac{2 R_E}{\omega \Delta t'} \frac{1}{\sqrt{1-\frac{p}{R_S}^2}}
\end{eqnarray}

Taking into account that Earth does not pass directly through the center of the shadow of Venus introduces a correction factor of 1.23899 to our original equation given the parameters of the transit on June 5/6, 2012 (with a value of $p = 0.5904 R_S$, as calculated by Backhaus \& Breil 2012).
 
Measuring the exact time of ingress exterior can be problematic because it is difficult to observe Venus before it begins to transit the Sun.  Therefore, rather than measure the time difference between first contact and second contact, we can measure the time difference between second contact as observed by two sets of observers at two different locations.  In doing so, note that the linear velocity of Earth ($v$, which we have assumed to be constant) is given by $2R_E/\Delta t$, which can be also be written as $\Delta x/\Delta t'$, where $x$ is the horizontal distance separating two points as the shadow moves across them.  This allows us to rewrite our full expression as:
\begin{eqnarray}
D = \frac{\Delta x}{\omega \Delta t'} \frac{1}{\sqrt{1-\frac{p}{R_S}^2}}
\end{eqnarray}

Unfortunately, this is not just a straightforward comparison of the latitude and longitude.  Due to a number of factors, including the position of the Sun and Venus on the sky (which depend on the date and time) and the tilt of EarthÕs axis, the shadow of Venus moves across the Earth at an angle relative to lines of constant latitude or longitude, as depicted in figure~5.  (For full details, see Backhaus \& Breil 2012).

\begin{figure}[htb]
\begin{center}
\includegraphics[width=8cm,angle=0]{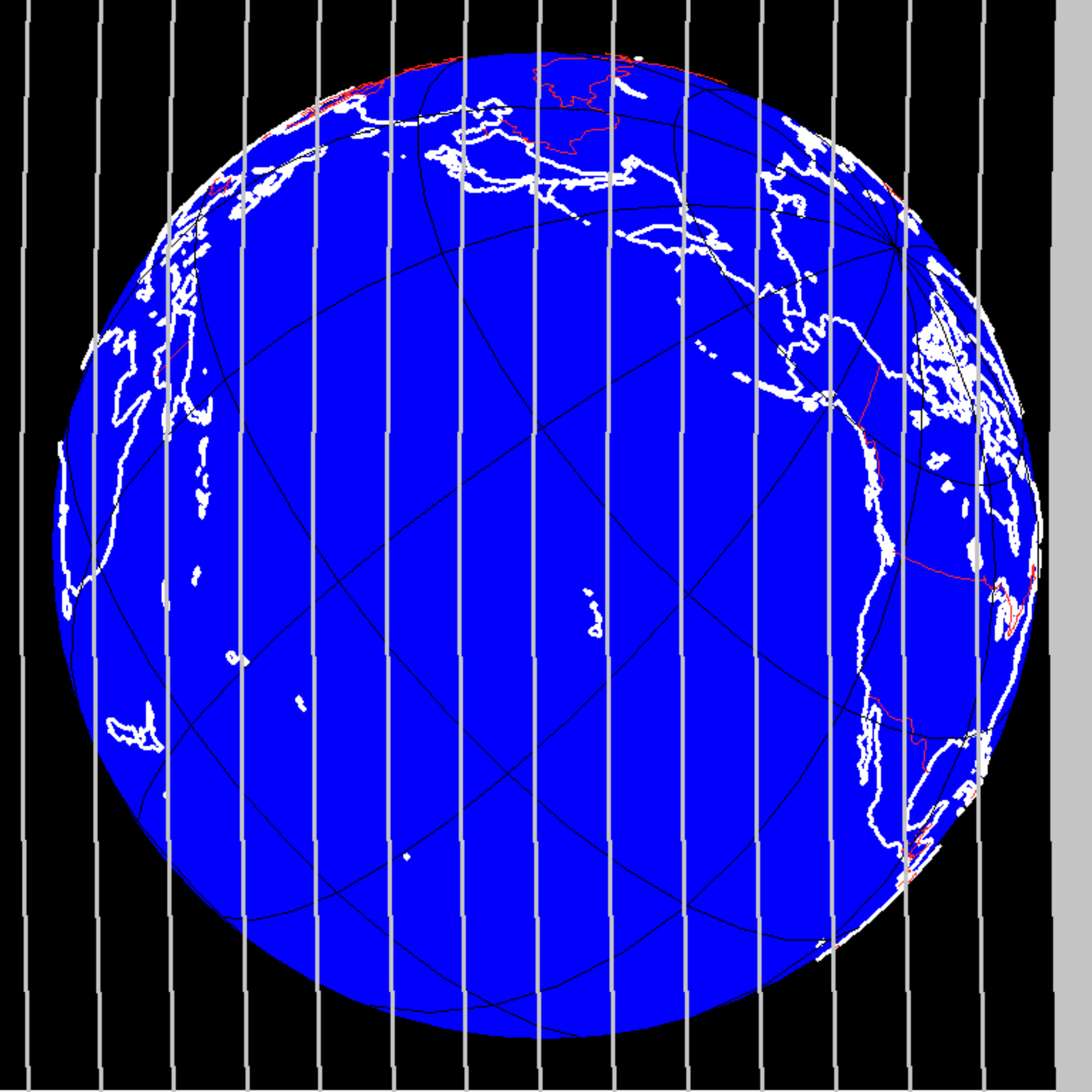}
\end{center}
\caption{The shadow of Venus moves from right to left in this diagram during the times of first and second contact.  The vertical grey lines correspond to values of constant $x$ for the 2012 transit of Venus.  (Credit:  Backhaus \& Breil 2012)
}
\end{figure}

Calculating $x$ given the latitude and longitude of each observer is straightforward, given the derivation:
\begin{eqnarray}
x =  0.4014 \cos \phi \cos \lambda - 0.5701 \cos \phi \sin \lambda + 0.7169 \sin \phi
\end{eqnarray}
where ($\phi$, $\lambda$) is the latitude and longitude of the observer, respectively. Note that this applies only for observers witnessing first or second contact. For those who see third or fourth contact, the relationship between latitude, longitude, and $x$ is:
\begin{eqnarray}
x =  -0.9371 \cos \phi \cos \lambda + 0.0438 \cos \phi \sin \lambda - 0.3463 \sin \phi
\end{eqnarray}

We now have a set of equations that allow us to calculate the distance from Earth to the Sun.  All that is needed is for at least two observers to record the time of second contact (or third contact) at each location, along with the latitude and longitude of each location, and the length of the Astronomical Unit can be determined. If we group all the constants together and express $\Delta t$ in seconds, we find that the distance between the Earth and the Sun is just:
\begin{eqnarray}
D = 6.35 \times 10^{10} \Delta x / \Delta t \, \text{(km)}
\end{eqnarray}


\section{Methodology}

In order to utilize the technique listed above, we required contact times from at least two locations separated by large distances on Earth. Furthermore, to ensure a successful measurement and to minimize problems caused by unpredictable weather, we recruited multiple teams strategically placed throughout the globe. Our initial team was formed by nine postdoctoral fellows working in mainland Chile. As our core group was observing from Easter Island, Chile, we dubbed our collaboration The Hetu'u Global Network, drawing from Rapanui, the native Polynesian language of Easter Island, and the word {\it hetu'u}, which means star. While other teams were working on similar projects (for example, a smartphone application which allowed users to time when a contact took place), we wanted to contact school and outreach groups in order to reach out to students and young children: the next generation of scientists. Consequently, after the core of our team was solidified, we began recruiting school groups from across the globe that would be observing the transit.  Initially the network was expanded using our personal outreach networks or known contacts observing the transit.  Eventually this was extended as we reached out to the organizers of the NOVA teacher sites, several Astronomy bloggers (including Universe Now), various NASA web-based lists, and the 2009 International Year of Astronomy participants.  This later listserv extended our reach to school groups in the far reaches of Iran, India, China, Colombia, and Japan.  

Through our efforts, we linked 19 school groups from the United States of America (New York, Florida, Texas, California, Hawaii), Australia, Japan, Hong Kong, Colombia, India, Iran, Holland, Denmark and Chile (see Figure 6). Given the diverse time zones and languages, we communicated with our primary contact people in English via email. Our team website also had additional information in both English and Spanish. The instructions were kept simple: \\
(1) Ensure that there is a GPS calibrated clock and that all participants at a given site are calibrated to the same time. \\
(2) Split each group into several smaller teams and have each observe the time of second and/or third contact (this was location dependent).  Multiple measurements at each site allowed us to gauge the uncertainty at each location.  \\
(3) Enter the GPS calibrated time of second and/or third contact as well as the longitude and latitude, and observing notes on the quality of data into a Google document which was shared among all participants.  Giving read/write access to each participant allowed each group to enter their own data and simultaneously have access to all other locations.  \\
(4) Take a photograph of the students observing the event since we were also interested in sharing the stories of the different groups who participated.

Emails were exchanged prior to and then throughout the time of the transit as groups kept each other updated on the location specific weather and quality of the data obtained.  

\section{Measurements}

Each group participating in our network had slightly different equipment, student numbers, and observing conditions. At the minimum end of equipment, groups used eclipse viewing glasses or sunspotter telescopes that project the image of the Sun on a larger screen.  At the high end of equipment, groups used medium sized telescopes with solar filters that provided a clear view of the transit. The procedure followed by all groups was the same: teams made note of second and/or third contact points with GPS calibrated clocks; however the results at each site varied.  Inclement weather prevented several groups from seeing one or both contact times.  Additionally, making a precise measurement of the transit contacts is inherently difficult. Turbulence in the atmosphere as well as imprecise equipment introduce uncertainties on exactly when second or third contact take place. Regardless, the teams worked with their given conditions and by combining all sites that were able to establish a timing measurement we were able to measure the distance to the Sun.  The map in Figure~6 shows the location of all teams and marks those that produced timing measurements and those that were weathered out.

\begin{figure}[htb]
\begin{center}
\includegraphics[width=14cm,angle=0]{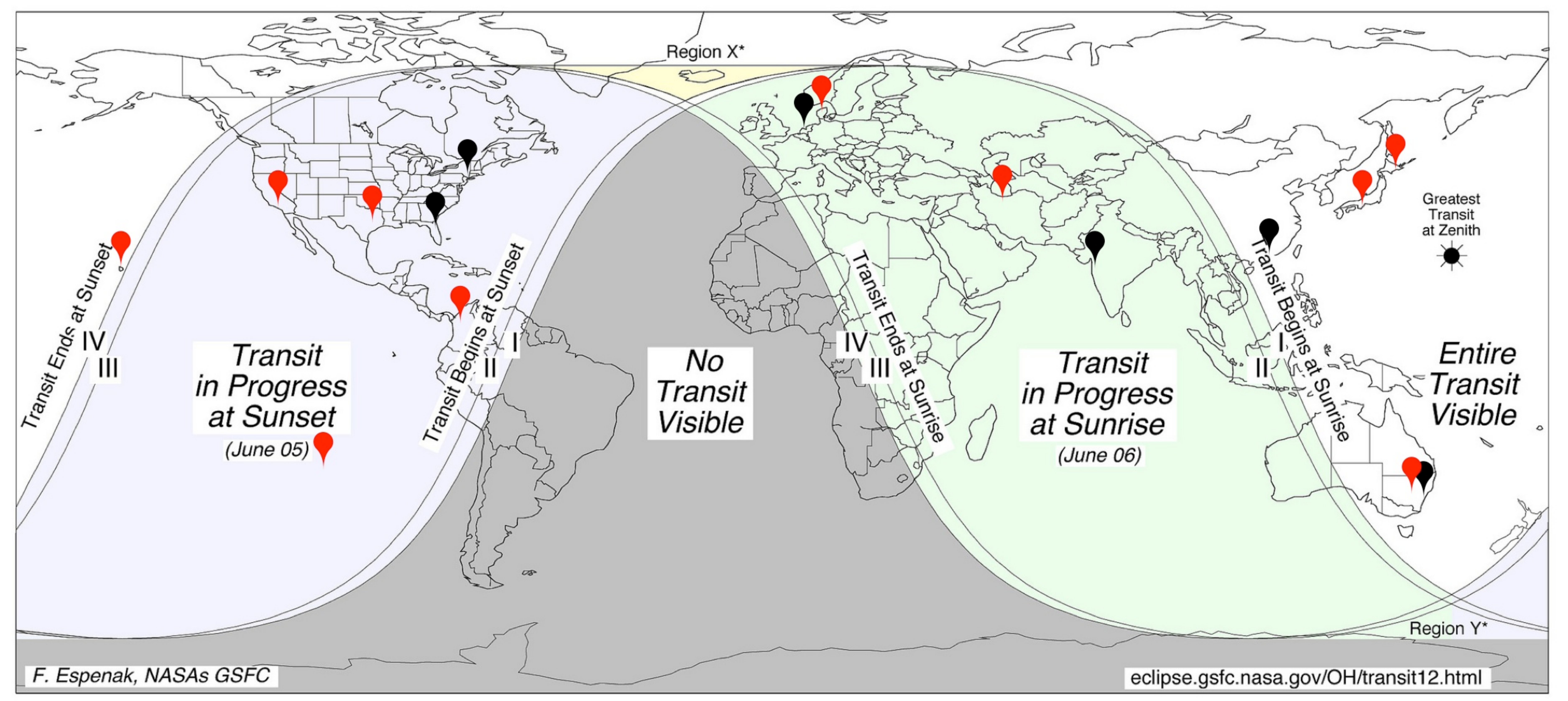}
\end{center}
\caption{Locations of school groups participating in a worldwide network to measure Earth's distance to the Sun using combined observations of the transit of Venus, overlaid on world visibility map for the transit of Venus (Credit: NASA).  The red indicators mark the locations of the groups that were able to record the time of second contact and/or third contact.
}
\end{figure}

\clearpage

\section{Results}

Among the 19 school groups participating in our global network, eight were able to estimate the important second Venus-Sun contact time and five were able to estimate the third Venus-Sun contact time.  Pictures of all participating groups can be found here: {\url http://www.flickr.com/photos/jfaherty/sets/72157630081603138/}.  Table~1 includes the average contact times recorded by each group that was successful in observing either one or both of the internal contact times, along with the groupsÕ latitudes and longitudes.  To calculate the distance to the sun we converted the latitude and longitude at each location into an $x$-Earth position using equation~7 for second contact and equation~8 for third contact.  We then converted the timing measurement at each location to seconds past either 22:00 GMT for second and 04:00 GMT for third Venus-Sun contact.  We note that several locations reported their contact times using their local clock time and two locations (Iran and India) are on half-hour time zones that required a correction.  Finally we used equation 9 to determine the Sun-Earth distance.  The results for both contact times are plotted in Figures~7a-7b.  We found distances of $152\pm30$ million km from second contact and $163\pm30$ million km from third contact measurements.  The uncertainty quoted for our measured distance ($\pm30$ million km) is estimated via bootstrapping. This is a statistical method in which the original, independent data points are used to generate new sets of data, thus allowing one to obtain the variance of some parameter. These new sets are generated by random sampling with replacement from the original dataset. Ten thousand such samples were generated in order to measure the uncertainty in our derived parameter: the distance between the Earth and the Sun.  The known distance to the Sun for that date was 152.25 million km.  Therefore, given our uncertainties of 30 million km, our calculated measurement is consistent with the true distance.

\begin{table}[htb]
\begin{tabular}{lrrrr}
\hline
Location & Latitude & Longitude & Second Contact & Third Contact  \\
& & & (GMT) & (GMT) \\
\hline
\hline
Huntsville, Texas, USA & 30.71314 & $-95.54818$ & 22:21:57 & \\
Los Angeles, California, USA & 34.0722 & $-118.44315$ & 22:23:15 & \\
Antioquia, Colombia & 8.62083 & $-76.93667$	 & 22:24:00 & \\
Paia, Hawaii, USA & 20.91652 & $-156.3847$ & 22:25:44 & 4:28:19 \\
Hyogo, Japan & 34.80013 & 134.8419 & 22:27:59 & 4:31:49 \\
Kyoto, Japan & 35.01167 & 135.76833 & 22:28:10 & 4:30:20 \\
Rey City (near Tehran), Iran & 35.6154 & 51.44508 & & 4:37:02 \\
Bjerget Efterskole, Denmark & 57.11139 & 8.99944 & & 4:38:00 \\
Easter Island, Chile & $-27.1557$ & $-109.42932$ & 22:28:10 & \\
Parkes, Australia & $-33.12639$ & 48.17417 & 22:34:40 & \\
\hline
\end{tabular}
\caption{Hetu'u Global Network contact time measurements}
\end{table}

\begin{figure}[htb]
\begin{center}
\includegraphics[width=12cm,angle=0]{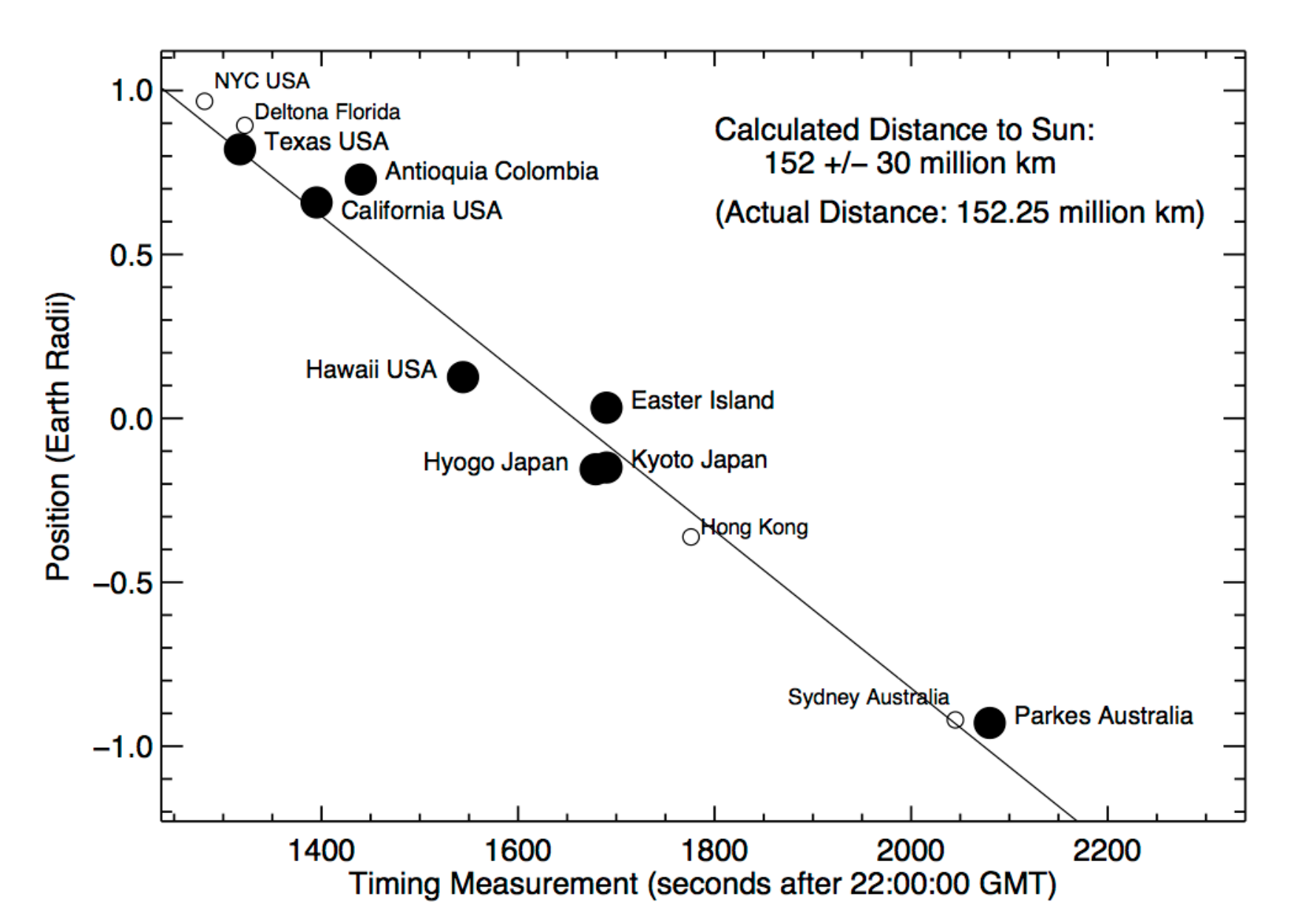}
\includegraphics[width=12cm,angle=0]{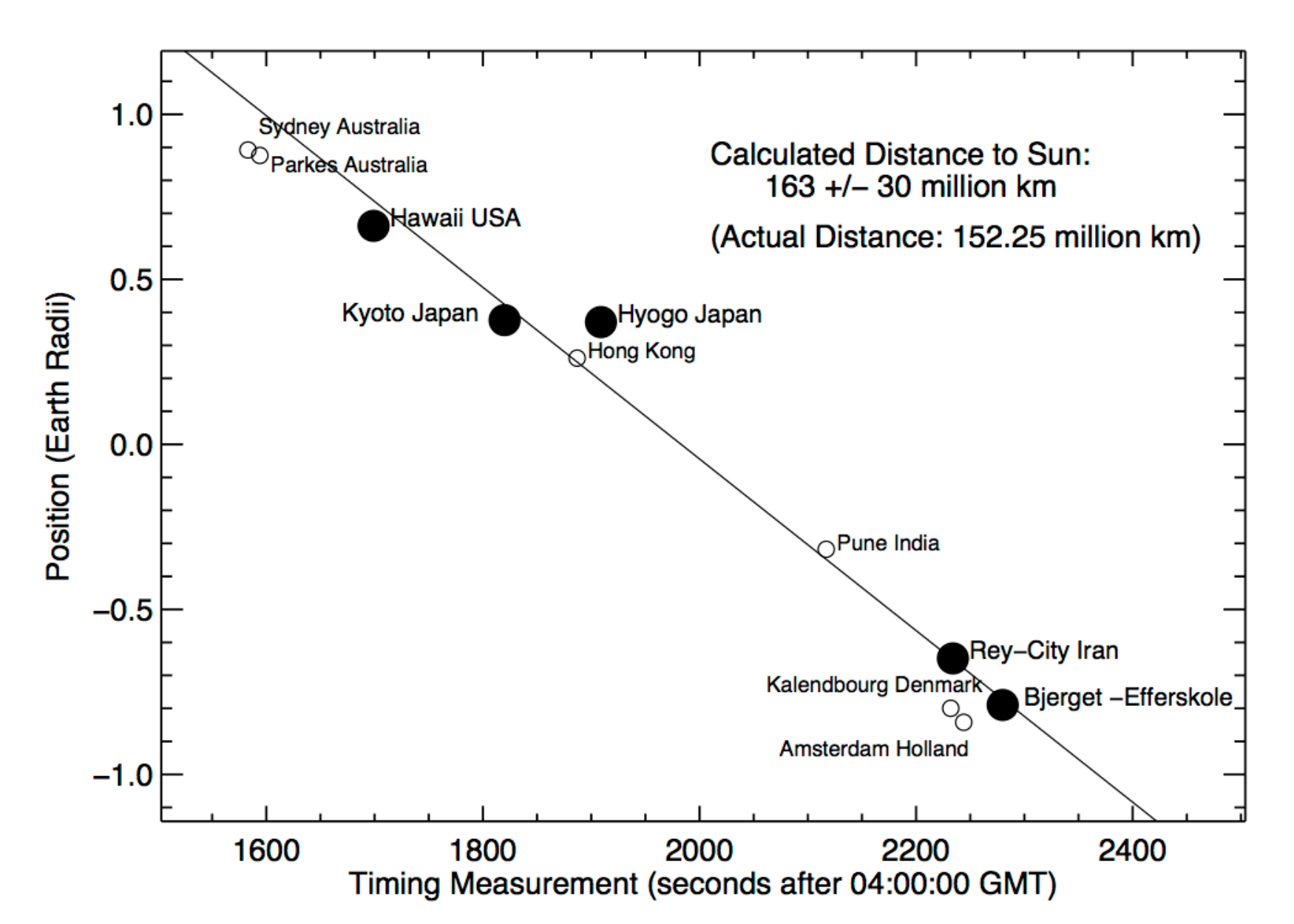}
\end{center}
\caption{Timing measurements of second (top) and third (bottom) Venus-Sun contacts taken around the globe (filled circles) along with the values expected for participating school groups where weather prevented viewings (open circles) as derived from van Roode \& Mignard (2012).   The best fit line to the data and calculated distance to the sun with uncertainty using the respective contact times are displayed.
}
\end{figure}

\clearpage

\section{Discussion}

Our distance measurements of $152\pm30$ and $163\pm30$ million km from second and third contacts respectively are consistent with the known distance to the Sun of 152.25 million km.  Given that our measurements were taken with amateur-grade telescopes and by a variety of observers (most of them high school or university students who have never observed before), the accuracy of our results is quite impressive.

Our greatest source of error results from the accuracy with which the various groups measured the time of ingress/egress interior.  Groups were instructed to separate into subgroups and have each make independent measurements to calculate a group average and standard deviation. Not all were able to do this; however, one group recorded eight distinct measurements for egress external, and based on their measurements, calculated a standard deviation of 16 seconds.  According to the formula to calculate the length of the astronomical unit, the distance to the Sun is equal to $6.35 \times 10^{10} \Delta x/\Delta t$. A one-second error in the time difference measurement taken in Parkes, Australia when compared to Texas, US (the two most widely separated measurements), would translate into an error of 0.19 million km.

The Black Drop effect, an optical phenomenon that is known to limit these measurements (see Schneider, Pasachoff, \& Golub 2004; Pasachoff, Schneider, \& Golub 2005; Pasachoff 2012), was not directly observed, but would have also added to our uncertainty.

\section{Conclusions}

For the June 5th/6th, 2012 transit of Venus, we gathered together 19 school groups from across the world to measure the distance between the Earth and the Sun using contact times. Ten such groups were able to measure second or third contact times leading to consistent measurements of the solar distance of $152\pm30$ and $163\pm30$ million km, respectively. Within uncertainties, these values are also consistent with the known distance to the Sun at the time of the transit. 

Our main goal, however, was to inspire the next generation of scientists using the excitement and accessibility of a rare astronomical event. Our Hetu'u Global Network of school groups connected hundreds of students from 6 continents and 10 countries representing a diverse, multi-cultural group with differing political, economic, and racial backgrounds.   

\clearpage

\acknowledgements
{\it Acknowledgements.} 
Participants in the Hetu'u Global Network were based all across the globe.  Below is a list of the city and country along with the primary contact and name of the school group (where available):

\noindent Hanga Roa on Easter Island, Chile\\
Colegio San Sebasti‡n de Akivi\\
Colegio Cat—lico Hermano Eugenio Eyraud\\
Colegio Lorenzo Baeza Vega\\
Contact: Jacqueline K. Faherty and David R.\ Rodriguez\\
Team Hetu'u Members: from Universidad de Chile: Jacqueline K. Faherty, David R.\ Rodriguez, Francisco F\"oster, Helene Flohic, Santiago Gonzalez, from Universidad Andr\'es Bello: Milena Bufano, Isabelle Gavignaud, from Universidad Cat\'olica: David Murphy, from Cerro Tololo: Catherine Kaleida; also present: Patricio Rojo and Natalie Huerta

\noindent Hawaii on the island of Maui\\
Students from the Pa'ia Youth \& Cultural Center of Maui (PYCC) \\
Contact: Adam Burgasser

\noindent Hawaii on the island of Maui\\
Students at the Haleakala Summit\\
Contact: James Armstrong

\noindent New York, New York USA\\
American Museum of Natural History students in the NASA Science Research Mentoring Program (SRMP) \\
Contact: Brian Levine

\noindent Los Angeles, California USA\\
Students gathered by the University of California Los Angeles (UCLA) astronomy outreach group (Astronomy Live!) \\
Contact: Breann Sitarski

\noindent Parkes, Australia\\
Parkes Observatory\\
Students from Sam Houston State University \\
Contact: Scott Miller and Renee James 

\noindent Sydney, Australia\\
Students from Sydney University \\
Contact: Joe Callingham

\clearpage
\noindent Huntsville, Texas\\
Students from Sam Houston State University\\
Contact: Mike Prokosch

\noindent Amsterdam, Holland\\
Students from Astronomical Institute Anton Pannekoek\\
Contact: Jaap vreeling 

\noindent Deltona, Florida USA\\
Students from Pine Ridge High School\\
Contact: Diane Sartore

\noindent Tehran, Iran\\
Students from Zakaria Razi Student Research Center  \\
Contact: Behnoosh Meskoob

\noindent Hong Kong, China\\
Students from Hong Kong Polytech  \\
Contact: John Babson

\noindent Hyogo, Japan\\
Kakogawa-Shi Sh\^onen Shizen no Ie\\
Contact: Alexandre Bouquin

\noindent Kyoto, Japan\\
Rakuy\^o Technical High School\\
Contact: Alexandre Bouquin

\noindent Kyoto, Japan\\
Kwazan Observatory, Kyoto University\\
Contact: Hiroki Kurokawa

\noindent Bjerget Efterskole, Denmark\\
Students from TINGH\O J\\
Contact: Poul Erik Groenhoej

\noindent Antioquia, Colombia\\
Universidad de Antioquia\\
Contact: Valentina Gonzalez

\noindent India\\
CSIR-URDIP\\
Contact: Vivek Doulatani

\noindent Kalendbourg, Denmark\\
Contact: Frank Wiehe Knudsen

\section*{References}

Backhaus, U. \& Breil, S. 2012, \\ {\url http://www.venus2012.de/venusprojects/contacttimes/details/detailstimes.php}

Forbes, G. 1874, ``The Transit of Venus," Macmillan and Co., London, 5

Pasachoff, Jay M., Glenn Schneider, \& Leon Golub, 2005, ``The black-drop effect explained," in Transits of Venus: New Views of the Solar System and Galaxy, IAU Colloquium No. 196 (U.K., 2004), D. W. Kurtz and G. E. Bromage, eds., 242-253.

Pasachoff, Jay M., 2012, ``Transit of Venus: Last Chance From Earth until 2117," Physics World, 25 (5), 36-41.

Schaefer, B. E. 2001, ``The transit of Venus and the notorious black drop effect," Journal for the History of Astronomy, 32(4), 109, 325

Schneider, Glenn, Jay M. Pasachoff, and Leon Golub 2004, ``TRACE Observations of the 15 November 1999 Transit of Mercury and the Black Drop Effect: Considerations for the 2004 Transit of Venus," Icarus 168, 249-256.

Short, J. 1761, ``The observations of the internal contact of Venus with the SunÕs limb, in the late transit, made in different places of Europe, compared with the time of the same contact observed at the Cape of Good Hope, and the parallax of the Sun from thence determined,"  Philosophical Transactions of the Royal Society, 52, 611

Teets, D. 2003, ``Transits of Venus and the Astronomical Unit," Mathematics Magazine, 76, 5

Van Roode, S. \& Mignard, F. 2012, {\url http://transitofvenus.nl/wp/where-when/local-transit-times/}

\clearpage

\end{document}